\documentclass{webofc}
\usepackage[varg]{txfonts}   
\usepackage{natbib}
\usepackage[dvipsnames]{xcolor}

\usepackage{fancyvrb}

\begin{document}

\title{AwkwardForth: accelerating Uproot with an internal DSL}

\author{\firstname{Jim} \lastname{Pivarski}\inst{1}\fnsep\thanks{\email{pivarski@princeton.edu}} \and
        \firstname{Ianna} \lastname{Osborne}\inst{1}\fnsep\thanks{\email{ianna.osborne@cern.ch}} \and
        \firstname{Pratyush} \lastname{Das}\inst{2}\fnsep\thanks{\email{reikdas@gmail.com}} \and
        \firstname{David} \lastname{Lange}\inst{1}\fnsep\thanks{\email{david.lange@cern.ch}} \and
        \firstname{Peter} \lastname{Elmer}\inst{1}\fnsep\thanks{\email{peter.elmer@cern.ch}}}

\institute{Princeton University, Princeton, New Jersey, United States. \and Institute of Engineering and Management, Kolkata, West Bengal, India.}

\abstract{
File formats for generic data structures, such as ROOT, Avro, and Parquet, pose a problem for deserialization: it must be fast, but its code depends on the type of the data structure, not known at compile-time. Just-in-time compilation can satisfy both constraints, but we propose a more portable solution: specialized virtual machines. AwkwardForth is a Forth-driven virtual machine for deserializing data into Awkward Arrays. As a language, it is not intended for humans to write, but it loosens the coupling between Uproot and Awkward Array. AwkwardForth programs for deserializing record-oriented formats (ROOT and Avro) are about as fast as C++ ROOT and 10--80$\times$ faster than fastavro. Columnar formats (simple TTrees, RNTuple, and Parquet) only require specialization to interpret metadata and are therefore faster with precompiled code.
}

\maketitle

\section{Motivation}

Despite being written in Python, Uproot~\cite{uproot} can read simple data types from ROOT TTrees~\cite{root} as fast as any precompiled code because the values are already contiguous in a raw view of the file. No computations are required to get a ROOT TBasket of numerical data from disk into memory, as an array, except to swap endianness if little-endian arrays are desired. Python's NumPy~\cite{numpy} library casts data from raw bytes as a metadata-only operation, and performs operations that scale as $\mathcal{O}(n)$, where $n$ is the length of the array, in precompiled code.

For more complex data types, however, the cost of computing in Python increases. Variable-length (``jagged'') arrays are relatively quick, since ROOT's TTree format separates the numerical content of these arrays from the integers that define the starting positions of each entry's array. This is nearly the format required by Awkward Array~\cite{awkward}, and only needs minor arithmetic transformations. Objects with fixed-size headers and consisting of fixed-size fields, such as {\tt std::vector<TLorentzVector>}, can be extracted using NumPy tricks, though these tricks require more intermediate arrays, further slowing the transformation. Finally, general objects with nested, variable-length data, the simplest of which is {\tt std::vector<std::vector<\textcolor{Maroon}{\textbf{float}}>{}>}, require custom code to parse each data type. In Python, such code is hundreds of times slower than the equivalent C++.

Previously, we quantified this slow-down~\cite{chep2019} using ROOT TTrees containing \textcolor{Maroon}{\tt\textbf{float}}, {\tt std::vector<\textcolor{Maroon}{\textbf{float}}>}, and vectors of vectors up to three levels deep, reading them with the then-current Python codebase and with custom C++ code, which represents what is possible now that Awkward Array is implemented in C++. The read performance of \textcolor{Maroon}{\tt\textbf{float}} data is identical for Python and C++, the C++ is several times faster for {\tt std::vector<\textcolor{Maroon}{\textbf{float}}>}, and the gap widens to factors of hundreds for the doubly-nested and triply-nested cases. Leveraging Awkward Array's C++ layer to accelerate Uproot is a worthwhile goal.

However, unless we limit our attention to special cases like nested vectors of numbers, this deserialization code is not known at compile-time. The byte-for-byte layout of a complex data type is expressed as data, in the TStreamerInfo of a ROOT file, and is therefore discovered at runtime. Moreover, the specifics of ROOT deserialization should not be spread between two packages, Uproot and Awkward Array: Uproot should focus entirely on ROOT I/O as Awkward Array focuses on array manipulation. The problem is to satisfy three constraints:
\begin{enumerate}
\item Deserialization code must not be hundreds of times slower than compiled code.
\item This code must be generated at runtime from TStreamerInfo.
\item All ``knowledge'' of ROOT I/O must be in Uproot, not Awkward Array.
\end{enumerate}

The first two constraints could be satisfied by just-in-time (JIT) compilation. This is, in fact, what ROOT's Cling compiler~\cite{cling} does. The third constraint could be satisfied by adding a C++ layer to Uproot. Both of these solutions, however, would significantly complicate the distribution of the Uproot or Awkward Array packages, and portability is a high priority.

This paper describes a different solution to the constraints listed above, which does not affect the portability of Uproot or Awkward Array. We introduce AwkwardForth, a domain-specific language (DSL) for deserializing arbitrary data into Awkward Arrays that is runtime-interpreted but nearly as fast as compiled code. Unlike DSLs intended for humans to read and write, this DSL is ``internal,'' only used to communicate between software packages. Uproot's task becomes one of expressing ROOT I/O logic in AwkwardForth and Awkward Array executes it, returning filled arrays.

\section{AwkwardForth}

AwkwardForth is a member of the Forth family of languages, which includes Postscript~\cite{postscript}, another internal DSL. AwkwardForth adheres to a subset of the ANSI Forth Standard~\cite{forth-standard} and has extensions for interpreting arbitrary input buffers and filling columns for Awkward Array.

Like all Forths, AwkwardForth is primarily concerned with stack manipulation. The runtime environment features a stack of integers and programs define ``words'' that manipulate the stack (as well as input and output buffers, in the case of AwkwardForth). Each word consumes and produces an arbitrary number of arguments and return values through this stack, and hence Forth words are more general than functions in a typical programming language. Whereas functional programming languages eliminate or minimize side-effects, Forth acts exclusively through side-effects. As such, it is more like an extensible assembly language.

The popularity of Forth peaked in the early 1980's because its interactive interpreter could fit within the tight resource constraints of early personal computers~\cite{forth}. This same interactive minimalism makes it an ideal candidate for running fast programs that must be generated at runtime, such as deserializing ROOT files. On typical hardware, AwkwardForth takes about 5~ns to evaluate each word, compared to about 900~ns for Python to evaluate a bytecode.

Though motivated by the problem of ROOT deserialization, AwkwardForth is intended for the general problem of deserializing non-columnar data formats into columnar Awkward Arrays. Many file formats, such as ProtoBuf~\cite{protobuf}, Thrift~\cite{thrift}, Avro~\cite{avro}, FlatBuffers~\cite{flatbuffers}, and JSON~\cite{json}, describe data structures in a record-oriented layout, with all fields of one record contiguous with each other, while columnar formats like simple TTrees, RNTuple~\cite{rntuple}, Parquet~\cite{parquet}, Arrow~\cite{arrow}, and Awkward Array place all values of one field contiguous with each other before moving on to the next field. Conversions between columnar formats can be very fast, sometimes casting, rather than copying, the columns. Record-oriented formats, on the other hand, must be fully parsed. In this paper, we examine AwkwardForth's deserialization performance for ROOT TTrees (columnar and record-oriented), Avro, and Parquet.

Awkward Array has a tool for converting arbitrary record-oriented data into Awkward Arrays: {\tt ArrayBuilder} constructs arrays in an append-only order, driven by commands such as {\tt begin}/{\tt end\_record}, switch to {\tt field}, append {\tt integer}, and {\tt begin}/{\tt end\_list}. A consequence of this generality is that {\tt ArrayBuilder} discovers the array's data type at runtime, adding output buffers as the observed type gets more complex. While this is great for JSON, type discovery is unnecessarily slow for formats whose type is known in advance, though perhaps not as early as compile-time. (Uproot currently uses {\tt ArrayBuilder} in Python.) In this paper, we also consider the design of a {\tt TypedArrayBuilder}, which generates AwkwardForth programs from data types, but is still driven by {\tt ArrayBuilder}-like commands.

\section{AwkwardForth virtual machine}

AwkwardForth is implemented in C++ as a virtual machine with byte-compiled instructions. It is not interactive, unlike most Forths, since it is intended to be programmed algorithmically. Even the {\tt TypedArrayBuilder} use-case would operate by ``wiring'' its fixed suite of commands to algorithmically generated words. Some {\tt TypedArrayBuilder} commands must change the state of its finite-state machine: for instance, when filling an array of doubly nested lists of integers, the first {\tt begin\_list} puts it into a state that expects another {\tt begin\_list} or {\tt end\_list}; the second puts it into a state that expects {\tt integer} or {\tt end\_list}. And yet, each of these commands must return control-flow to its caller and remember its state for the next call. To implement {\tt TypedArrayBuilder} with an AwkwardForth machine, the AwkwardForth machine must be able to stop and resume with its state intact (unlike most Forths). AwkwardForth, therefore, has built-in words to control its own execution (\textcolor{OliveGreen}{\tt\textbf{pause}} and \textcolor{OliveGreen}{\tt\textbf{halt}}), and the execution may be resumed from outside the machine.

When an AwkwardForth machine is first constructed, it compiles its source code (text) into bytecode instructions (variable-length sequences of 32-bit integers---a jagged array), so that so that they can be interpreted more quickly. This is the same sense in which Python code is ``compiled.'' Built-in words translate to 1--3 integer codes, the second and third being arguments that modify the first. User-defined words are separate sequences of instructions, called a ``dictionary'' in Forth. Control-flow structures, such as \textcolor{OliveGreen}{\tt\textbf{if}} and \textcolor{OliveGreen}{\tt\textbf{loop}}, are implemented as unnamed user-defined words so that their bodies are fixed-width ``calls'' into the dictionary, simplifying the logic of instruction pointer manipulation.

All errors are caught in the compilation phase except for 10 possible runtime errors: ``user halt,'' ``recursion depth exceeded,'' ``stack underflow,'' ``stack overflow,'' ``division by zero,'' ``read beyond,'' ``seek beyond,'' ``skip beyond,'' ``rewind beyond,'' and ``varint too big.'' The last 5 of these are specific to parsing input buffers.

All runtime execution is implemented in a single ``\textcolor{OliveGreen}{\tt\textbf{noexcept}}'' C++ function for speed.

An arbitrary number of named input buffers and named output buffers are associated with each AwkwardForth program. Inputs must be supplied at the beginning of a run; outputs are also created at this time. Input buffers, which are the data to be parsed, are viewed as untyped raw bytes, interpreted by the words of the program itself. Output buffers, which are columns to use in a new Awkward Array, have specific numerical types and can only be filled with that type. Inputs have a fixed size and are seekable; outputs grow in an append-only way.

\section{AwkwardForth language}

Like all Forth languages, AwkwardForth has an extremely simple syntax: whitespace delimits tokens and tokens are interpreted in reverse Polish order: ``{\tt\textcolor{gray}{3} \textcolor{gray}{4} \textcolor{OliveGreen}{\textbf{+}}}'' means the sum of \textcolor{gray}{\tt 3} and \textcolor{gray}{\tt 4} (the first two words put numbers on the stack and the third pops them and pushes their sum).

AwkwardForth has 57 standard built-in words and 61 extensions for input and output buffers, though 50 of the 61 are different ways of interpreting the bytes of an input buffer. Two special words control the state of the virtual machine: \textcolor{OliveGreen}{\tt\textbf{pause}} and \textcolor{OliveGreen}{\tt\textbf{halt}}, and three do not generate runtime code at all: they declare variables, inputs, and outputs.

\small
\begin{Verbatim}[commandchars=\\\{\}]
   \textcolor{OliveGreen}{\textbf{variable}} \textcolor{blue}{\textbf{variable_name}}               \textcolor{teal}{\textit{( Standard Forth )}}
   \textcolor{OliveGreen}{\textbf{input   }} \textcolor{blue}{\textbf{input_name}}                  \textcolor{teal}{\textit{( AwkwardForth extension )}}
   \textcolor{OliveGreen}{\textbf{output  }} \textcolor{blue}{\textbf{output_name}} output_type     \textcolor{teal}{\textit{( AwkwardForth extension )}}
\end{Verbatim}
\normalsize

As in Standard Forth, user-defined words are bracketed between \textcolor{OliveGreen}{\tt\textbf{:}} and \textcolor{OliveGreen}{\tt\textbf{;}} and the main code is anything outside these definitions. Control structures are pairs of words like \textcolor{OliveGreen}{\tt\textbf{if}}-\textcolor{OliveGreen}{\tt\textbf{then}} and \textcolor{OliveGreen}{\tt\textbf{do}}-\textcolor{OliveGreen}{\tt\textbf{loop}}, or triples like \textcolor{OliveGreen}{\tt\textbf{if}}-\textcolor{OliveGreen}{\tt\textbf{else}}-\textcolor{OliveGreen}{\tt\textbf{then}} and \textcolor{OliveGreen}{\tt\textbf{begin}}-\textcolor{OliveGreen}{\tt\textbf{while}}-\textcolor{OliveGreen}{\tt\textbf{repeat}}.

\small
\begin{Verbatim}[commandchars=\\\{\}]
   \textcolor{OliveGreen}{\textbf{:}} \textcolor{blue}{\textbf{fibonacci}}    \textcolor{teal}{\textit{( pops n -- pushes nth-fibonacci-number )}}
     \textcolor{OliveGreen}{\textbf{dup}}
     \textcolor{gray}{1} \textcolor{OliveGreen}{\textbf{>}} \textcolor{OliveGreen}{\textbf{if}}
       \textcolor{OliveGreen}{\textbf{1-}} \textcolor{OliveGreen}{\textbf{dup}} \textcolor{OliveGreen}{\textbf{1-}} \textcolor{blue}{fibonacci}
       \textcolor{OliveGreen}{\textbf{swap}} \textcolor{blue}{fibonacci}
       \textcolor{OliveGreen}{\textbf{+}}
     \textcolor{OliveGreen}{\textbf{then}}
   \textcolor{OliveGreen}{\textbf{;}}

   \textcolor{teal}{\textit{( pushes [0 1 1 2 3 5 8 13 21 34 55 89 144 233 377] onto the stack )}}
   \textcolor{gray}{15} \textcolor{gray}{0} \textcolor{OliveGreen}{\textbf{do}}
     \textcolor{OliveGreen}{\textbf{i}} \textcolor{blue}{fibonacci}
   \textcolor{OliveGreen}{\textbf{loop}}
\end{Verbatim}
\normalsize

The 50 special words for parsing are all arrows with a type code, an optional \textcolor{OliveGreen}{\tt\textbf{!}} for big-endian and an optional \textcolor{OliveGreen}{\tt\textbf{\#}} to pull a number off the stack to determine how many times to repeat it. Thus, ``{\tt \textcolor{blue}{input\_name} \textcolor{OliveGreen}{\textbf{i->}} \textcolor{blue}{output\_name}}'' reads 4 bytes from the input as an integer and writes it to the output, ``{\tt \textcolor{blue}{input\_name} \textcolor{OliveGreen}{\textbf{!d->}} \textcolor{blue}{output\_name}}'' reads 8 bytes as a big-endian, double-precision float, and ``{\tt \textcolor{gray}{100} \textcolor{blue}{input\_name} \textcolor{OliveGreen}{\textbf{\#I->}} \textcolor{blue}{output\_name}}'' reads 100 unsigned 4-byte integers. In any case, the destination may be the stack: ``{\tt \textcolor{blue}{input\_name} \textcolor{OliveGreen}{\textbf{i->}} \textcolor{OliveGreen}{\textbf{stack}}}''.

The type codes are taken from Python's {\tt struct} module, which AwkwardForth most closely resembles in purpose (but vastly exceeds in expressiveness). Two special codes, \textcolor{OliveGreen}{\tt\textbf{varint->}} and \textcolor{OliveGreen}{\tt\textbf{zigzag->}}, read variable-length unsigned integers and zig-zag encoded signed integers, which are used in many file formats, including Avro and Parquet. Also, Parquet needs a command to read $n$-bit unsigned integers, hence \textcolor{OliveGreen}{\tt\textbf{2bit->}}, \textcolor{OliveGreen}{\tt\textbf{3bit->}}, etc.\ for any $n$.

Outputs are similar, but less varied because they have preassigned types. They may be filled directly from an input (bypassing the stack for efficiency and to preserve the types of floating point numbers) or filled from the stack: ``{\tt \textcolor{blue}{output\_name} \textcolor{OliveGreen}{\textbf{<-}} \textcolor{OliveGreen}{\textbf{stack}}}''. A shortcut for appending the last output plus a value from the stack is ``{\tt \textcolor{blue}{output\_name} \textcolor{OliveGreen}{\textbf{+<-}} \textcolor{OliveGreen}{\textbf{stack}}}''.

More built-in words can be added to handle problems posed by input formats. Since each built-in word has a fixed 5~ns cost, specialized words result in faster AwkwardForth code.

\section{AwkwardForth programs for ROOT, Avro, and Parquet}
\label{sec:programs}

To read a TBasket of {\tt std::vector<std::vector<\textcolor{Maroon}{\textbf{float}}>{}>} from ROOT, we use the following AwkwardForth program:

\small
\begin{Verbatim}[commandchars=\\\{\}]
   \textcolor{OliveGreen}{\textbf{input}} \textcolor{blue}{\textbf{data}}                      \textcolor{teal}{\textit{( ROOT TBaskets have a data buffer )}}
   \textcolor{OliveGreen}{\textbf{input}} \textcolor{blue}{\textbf{byte_offsets}}              \textcolor{teal}{\textit{( and a buffer of byte offsets )}}

   \textcolor{OliveGreen}{\textbf{output}} \textcolor{blue}{\textbf{offsets0}} int32           \textcolor{teal}{\textit{( output Awkward Array offsets )}}
   \textcolor{OliveGreen}{\textbf{output}} \textcolor{blue}{\textbf{offsets1}} int32           \textcolor{teal}{\textit{( ... )}}
   \textcolor{OliveGreen}{\textbf{output}} \textcolor{blue}{\textbf{offsets2}} int32           \textcolor{teal}{\textit{( ... )}}
   \textcolor{OliveGreen}{\textbf{output}} \textcolor{blue}{\textbf{content}} float32          \textcolor{teal}{\textit{( and content )}}

   \textcolor{gray}{\textbf{0}} \textcolor{blue}{offsets0} \textcolor{OliveGreen}{\textbf{<-}} \textcolor{OliveGreen}{\textbf{stack}}             \textcolor{teal}{\textit{( offsets start at zero )}}
   \textcolor{gray}{\textbf{0}} \textcolor{blue}{offsets1} \textcolor{OliveGreen}{\textbf{<-}} \textcolor{OliveGreen}{\textbf{stack}}
   \textcolor{gray}{\textbf{0}} \textcolor{blue}{offsets2} \textcolor{OliveGreen}{\textbf{<-}} \textcolor{OliveGreen}{\textbf{stack}}

   \textcolor{OliveGreen}{\textbf{begin}}
     \textcolor{blue}{byte_offsets} \textcolor{OliveGreen}{\textbf{i->}} \textcolor{OliveGreen}{\textbf{stack}}        \textcolor{teal}{\textit{( get a position from the byte offsets )}}
     \textcolor{gray}{6} \textcolor{OliveGreen}{\textbf{+}} \textcolor{blue}{data} \textcolor{OliveGreen}{\textbf{seek}}                 \textcolor{teal}{\textit{( seek to it plus a 6-byte header )}}
     \textcolor{blue}{data} \textcolor{OliveGreen}{\textbf{!i->}} \textcolor{OliveGreen}{\textbf{stack}}               \textcolor{teal}{\textit{( get the std::vector size )}}
     \textcolor{OliveGreen}{\textbf{dup}} \textcolor{blue}{offsets0} \textcolor{OliveGreen}{\textbf{+<-}} \textcolor{OliveGreen}{\textbf{stack}}        \textcolor{teal}{\textit{( add it to the offsets )}}
     \textcolor{gray}{0} \textcolor{OliveGreen}{\textbf{do}}                          \textcolor{teal}{\textit{( and use it as the loop counter )}}
       \textcolor{blue}{data} \textcolor{OliveGreen}{\textbf{!i->}} \textcolor{OliveGreen}{\textbf{stack}}             \textcolor{teal}{\textit{( same for the inner std::vector )}}
       \textcolor{OliveGreen}{\textbf{dup}} \textcolor{blue}{offsets1} \textcolor{OliveGreen}{\textbf{+<-}} \textcolor{OliveGreen}{\textbf{stack}}
       \textcolor{gray}{0} \textcolor{OliveGreen}{\textbf{do}}
         \textcolor{blue}{data} \textcolor{OliveGreen}{\textbf{!i->}} \textcolor{OliveGreen}{\textbf{stack}}           \textcolor{teal}{\textit{( and the innermost std::vector )}}
         \textcolor{OliveGreen}{\textbf{dup}} \textcolor{blue}{offsets2} \textcolor{OliveGreen}{\textbf{+<-}} \textcolor{OliveGreen}{\textbf{stack}}
         \textcolor{blue}{data} \textcolor{OliveGreen}{\textbf{#!f->}} \textcolor{blue}{content}        \textcolor{teal}{\textit{( finally, the floating point values )}}
       \textcolor{OliveGreen}{\textbf{loop}}
      \textcolor{OliveGreen}{\textbf{loop}}
   \textcolor{OliveGreen}{\textbf{again}}                           \textcolor{teal}{\textit{( ends with a "seek beyond" exception )}}
\end{Verbatim}
\normalsize

\noindent To read data with the same structure from Avro, we use the program below. It is applied to each data block of an Avro container file, with the AwkwardForth machine stack initialized with the number of entries in the block.

\small
\begin{Verbatim}[commandchars=\\\{\}]
   \textcolor{OliveGreen}{\textbf{input}} \textcolor{blue}{\textbf{data}}                      \textcolor{teal}{\textit{( an Avro data block is a single buffer )}}

   \textcolor{OliveGreen}{\textbf{output}} \textcolor{blue}{\textbf{offsets0}} int32           \textcolor{teal}{\textit{( output Awkward Array offsets )}}
   \textcolor{OliveGreen}{\textbf{output}} \textcolor{blue}{\textbf{offsets1}} int32           \textcolor{teal}{\textit{( ... )}}
   \textcolor{OliveGreen}{\textbf{output}} \textcolor{blue}{\textbf{offsets2}} int32           \textcolor{teal}{\textit{( ... )}}
   \textcolor{OliveGreen}{\textbf{output}} \textcolor{blue}{\textbf{content}} float32          \textcolor{teal}{\textit{( and content )}}

   \textcolor{gray}{\textbf{0}} \textcolor{blue}{offsets0} \textcolor{OliveGreen}{\textbf{<-}} \textcolor{OliveGreen}{\textbf{stack}}             \textcolor{teal}{\textit{( offsets start at zero )}}
   \textcolor{gray}{\textbf{0}} \textcolor{blue}{offsets1} \textcolor{OliveGreen}{\textbf{<-}} \textcolor{OliveGreen}{\textbf{stack}}
   \textcolor{gray}{\textbf{0}} \textcolor{blue}{offsets2} \textcolor{OliveGreen}{\textbf{<-}} \textcolor{OliveGreen}{\textbf{stack}}

   \textcolor{gray}{0} \textcolor{OliveGreen}{\textbf{do}}                            \textcolor{teal}{\textit{( upper limit on the stack at startup )}}
     \textcolor{blue}{data} \textcolor{OliveGreen}{\textbf{zigzag->}} \textcolor{OliveGreen}{\textbf{stack}}
     \textcolor{OliveGreen}{\textbf{dup}} \textcolor{blue}{offsets0} \textcolor{OliveGreen}{\textbf{+<-}} \textcolor{OliveGreen}{\textbf{stack}}        \textcolor{teal}{\textit{( add it to the offsets )}}
     \textcolor{gray}{0} \textcolor{OliveGreen}{\textbf{do}}                          \textcolor{teal}{\textit{( and use it as the loop counter )}}
       \textcolor{blue}{data} \textcolor{OliveGreen}{\textbf{zigzag->}} \textcolor{OliveGreen}{\textbf{stack}}
       \textcolor{OliveGreen}{\textbf{dup}} \textcolor{blue}{offsets1} \textcolor{OliveGreen}{\textbf{+<-}} \textcolor{OliveGreen}{\textbf{stack}}
       \textcolor{gray}{0} \textcolor{OliveGreen}{\textbf{do}}
         \textcolor{blue}{data} \textcolor{OliveGreen}{\textbf{zigzag->}} \textcolor{OliveGreen}{\textbf{stack}}
         \textcolor{OliveGreen}{\textbf{dup}} \textcolor{blue}{offsets2} \textcolor{OliveGreen}{\textbf{+<-}} \textcolor{OliveGreen}{\textbf{stack}}
         \textcolor{blue}{data} \textcolor{OliveGreen}{\textbf{#f->}} \textcolor{blue}{content}         \textcolor{teal}{\textit{( finally, the floating point values )}}
         \textcolor{blue}{data} \textcolor{OliveGreen}{\textbf{b->}} \textcolor{OliveGreen}{\textbf{stack}} \textcolor{OliveGreen}{\textbf{drop}}       \textcolor{teal}{\textit{( ends with a zero-byte )}}
       \textcolor{OliveGreen}{\textbf{loop}}
       \textcolor{blue}{data} \textcolor{OliveGreen}{\textbf{b->}} \textcolor{OliveGreen}{\textbf{stack}} \textcolor{OliveGreen}{\textbf{drop}}         \textcolor{teal}{\textit{( lists also end with a zero-byte )}}
     \textcolor{OliveGreen}{\textbf{loop}}
     \textcolor{blue}{data} \textcolor{OliveGreen}{\textbf{b->}} \textcolor{OliveGreen}{\textbf{stack}} \textcolor{OliveGreen}{\textbf{drop}}           \textcolor{teal}{\textit{( lists also end with a zero-byte )}}
   \textcolor{OliveGreen}{\textbf{loop}}
\end{Verbatim}
\normalsize

\noindent Parquet is a columnar file format, so the floating-point content comes in a form that's ready to use in Awkward Array, without even copying the uncompressed buffer. However, the nested list structure is in a highly packed form: repetition levels indicating the depth of each floating-point item, which are further run-length encoded or bit-packed in small groups.

We unpack them using two AwkwardForth programs: the first produces the repetition levels as unsigned 1-byte integers and the second converts them into three levels of offsets.

\small
\begin{Verbatim}[commandchars=\\\{\}]
   \textcolor{OliveGreen}{\textbf{input}} \textcolor{blue}{\textbf{data}}
   \textcolor{OliveGreen}{\textbf{output}} \textcolor{blue}{\textbf{replevels}} uint8          \textcolor{teal}{\textit{( output 1-byte integers )}}

   \textcolor{blue}{data} \textcolor{OliveGreen}{\textbf{I->}} \textcolor{OliveGreen}{\textbf{stack}}                  \textcolor{teal}{\textit{( get the number of bytes to read )}}
   \textcolor{OliveGreen}{\textbf{begin}}
     \textcolor{blue}{data} \textcolor{OliveGreen}{\textbf{varint->}} \textcolor{OliveGreen}{\textbf{stack}}           \textcolor{teal}{\textit{( read a variable-length integer )}}
     \textcolor{OliveGreen}{\textbf{dup}} \textcolor{gray}{1} \textcolor{OliveGreen}{\textbf{and}}                     \textcolor{teal}{\textit{( its lowest bit selects encoding type )}}

     \textcolor{OliveGreen}{\textbf{0=}} \textcolor{OliveGreen}{\textbf{if}}                         \textcolor{teal}{\textit{( zero means run-length encoding... )}}
       \textcolor{blue}{data} \textcolor{OliveGreen}{\textbf{B->}} \textcolor{blue}{replevels}          \textcolor{teal}{\textit{( write the value to duplicate )}}
       \textcolor{gray}{1} \textcolor{OliveGreen}{\textbf{rshift}} \textcolor{OliveGreen}{\textbf{1-}}                 \textcolor{teal}{\textit{( determine how many times )}}
       \textcolor{blue}{replevels} \textcolor{OliveGreen}{\textbf{dup}}               \textcolor{teal}{\textit{( duplicate it )}}

     \textcolor{OliveGreen}{\textbf{else}}                          \textcolor{teal}{\textit{( non-zero means bit-packed... )}}
       \textcolor{gray}{1} \textcolor{OliveGreen}{\textbf{rshift}} \textcolor{gray}{8} \textcolor{OliveGreen}{\textbf{*}}                \textcolor{teal}{\textit{( determine how many to read )}}
       \textcolor{blue}{data} \textcolor{OliveGreen}{\textbf{#2bit->}} \textcolor{blue}{replevels}      \textcolor{teal}{\textit{( read all the 2-bit integers )}}
     \textcolor{OliveGreen}{\textbf{then}}

     \textcolor{OliveGreen}{\textbf{dup}} \textcolor{blue}{data} \textcolor{OliveGreen}{\textbf{pos}} \textcolor{gray}{4} \textcolor{OliveGreen}{\textbf{-}}              \textcolor{teal}{\textit{( continue until end of input buffer )}}
   \textcolor{OliveGreen}{\textbf{until}}
\end{Verbatim}
\normalsize

\noindent This last program interprets the repetition levels as offsets. It uses variables to count the number of items at each level of list depth since all three need to be increased concurrently and using \textcolor{OliveGreen}{\tt\textbf{swap}} or \textcolor{OliveGreen}{\tt\textbf{rot}} to manage them on the stack would be unnecessarily complex. The Standard Forth words \textcolor{OliveGreen}{\tt\textbf{@}}, \textcolor{OliveGreen}{\tt\textbf{!}}, and \textcolor{OliveGreen}{\tt\textbf{+!}} read, write, and increment an off-stack variable. Variables have the same integer type as the stack and are initially zero.

\small
\begin{Verbatim}[commandchars=\\\{\}]
   \textcolor{OliveGreen}{\textbf{input}} \textcolor{blue}{\textbf{replevels}}
   \textcolor{OliveGreen}{\textbf{output}} \textcolor{blue}{\textbf{offsets0}} int32 \textcolor{OliveGreen}{\textbf{output}} \textcolor{blue}{\textbf{offsets1}} int32 \textcolor{OliveGreen}{\textbf{output}} \textcolor{blue}{\textbf{offsets2}} int32
   \textcolor{OliveGreen}{\textbf{variable}} \textcolor{blue}{\textbf{count0}} \textcolor{OliveGreen}{\textbf{variable}} \textcolor{blue}{\textbf{count1}} \textcolor{OliveGreen}{\textbf{variable}} \textcolor{blue}{\textbf{count2}}

   \textcolor{OliveGreen}{\textbf{begin}}
     \textcolor{blue}{replevels} \textcolor{OliveGreen}{\textbf{b->}} \textcolor{OliveGreen}{\textbf{stack}}           \textcolor{teal}{\textit{( get one repetition level )}}

     \textcolor{OliveGreen}{\textbf{dup}} \textcolor{gray}{3} \textcolor{OliveGreen}{\textbf{=}} \textcolor{OliveGreen}{\textbf{if}}                    \textcolor{teal}{\textit{( 3 means deepest level of structure )}}
       \textcolor{gray}{1} \textcolor{blue}{count2} \textcolor{OliveGreen}{\textbf{+!}}
     \textcolor{OliveGreen}{\textbf{then}}
     \textcolor{OliveGreen}{\textbf{dup}} \textcolor{gray}{2} \textcolor{OliveGreen}{\textbf{=}} \textcolor{OliveGreen}{\textbf{if}}                    \textcolor{teal}{\textit{( 2 means a new innermost list )}}
       \textcolor{gray}{1} \textcolor{blue}{count1} \textcolor{OliveGreen}{\textbf{+!}}
       \textcolor{blue}{count2} \textcolor{OliveGreen}{\textbf{@}} \textcolor{blue}{offsets2} \textcolor{OliveGreen}{\textbf{+<-}} \textcolor{OliveGreen}{\textbf{stack}} \textcolor{gray}{1} \textcolor{blue}{count2} \textcolor{OliveGreen}{\textbf{!}}
     \textcolor{OliveGreen}{\textbf{then}}
     \textcolor{OliveGreen}{\textbf{dup}} \textcolor{gray}{1} \textcolor{OliveGreen}{\textbf{=}} \textcolor{OliveGreen}{\textbf{if}}                    \textcolor{teal}{\textit{( 1 means a new inner list )}}
       \textcolor{gray}{1} \textcolor{blue}{count0} \textcolor{OliveGreen}{\textbf{+!}}
       \textcolor{blue}{count1} \textcolor{OliveGreen}{\textbf{@}} \textcolor{blue}{offsets1} \textcolor{OliveGreen}{\textbf{+<-}} \textcolor{OliveGreen}{\textbf{stack}} \textcolor{gray}{1} \textcolor{blue}{count1} \textcolor{OliveGreen}{\textbf{!}}
       \textcolor{blue}{count2} \textcolor{OliveGreen}{\textbf{@}} \textcolor{blue}{offsets2} \textcolor{OliveGreen}{\textbf{+<-}} \textcolor{OliveGreen}{\textbf{stack}} \textcolor{gray}{1} \textcolor{blue}{count2} \textcolor{OliveGreen}{\textbf{!}}
     \textcolor{OliveGreen}{\textbf{then}}
     \textcolor{gray}{0} \textcolor{OliveGreen}{\textbf{=}} \textcolor{OliveGreen}{\textbf{if}}                        \textcolor{teal}{\textit{( 0 means a new outer list )}}
       \textcolor{blue}{count0} \textcolor{OliveGreen}{\textbf{@}} \textcolor{blue}{offsets0} \textcolor{OliveGreen}{\textbf{+<-}} \textcolor{OliveGreen}{\textbf{stack}} \textcolor{gray}{1} \textcolor{blue}{count0} \textcolor{OliveGreen}{\textbf{!}}
       \textcolor{blue}{count1} \textcolor{OliveGreen}{\textbf{@}} \textcolor{blue}{offsets1} \textcolor{OliveGreen}{\textbf{+<-}} \textcolor{OliveGreen}{\textbf{stack}} \textcolor{gray}{1} \textcolor{blue}{count1} \textcolor{OliveGreen}{\textbf{!}}
       \textcolor{blue}{count2} \textcolor{OliveGreen}{\textbf{@}} \textcolor{blue}{offsets2} \textcolor{OliveGreen}{\textbf{+<-}} \textcolor{OliveGreen}{\textbf{stack}} \textcolor{gray}{1} \textcolor{blue}{count2} \textcolor{OliveGreen}{\textbf{!}}
     \textcolor{OliveGreen}{\textbf{then}}

     \textcolor{blue}{replevels} \textcolor{OliveGreen}{\textbf{end}}                 \textcolor{teal}{\textit{( continue to the end of input )}}
   \textcolor{OliveGreen}{\textbf{until}}

   \textcolor{blue}{count0} \textcolor{OliveGreen}{\textbf{@}} \textcolor{blue}{offsets0} \textcolor{OliveGreen}{\textbf{+<-}} \textcolor{OliveGreen}{\textbf{stack}}     \textcolor{teal}{\textit{( add the last counts for all three )}}
   \textcolor{blue}{count1} \textcolor{OliveGreen}{\textbf{@}} \textcolor{blue}{offsets1} \textcolor{OliveGreen}{\textbf{+<-}} \textcolor{OliveGreen}{\textbf{stack}}
   \textcolor{blue}{count2} \textcolor{OliveGreen}{\textbf{@}} \textcolor{blue}{offsets2} \textcolor{OliveGreen}{\textbf{+<-}} \textcolor{OliveGreen}{\textbf{stack}}
\end{Verbatim}
\normalsize

Programs such as these would not be written (and commented!) by hand, but generated by Uproot, other file-readers that produce Awkward Arrays, or {\tt TypedArrayBuilder}.

\section{Example of an AwkwardForth program for TypedArrayBuilder}

The {\tt TypedArrayBuilder} is still in development, but the following illustrates AwkwardForth that can be generated for a triply nested list of \textcolor{Maroon}{\tt\textbf{float}}.

\small
\begin{Verbatim}[commandchars=\\\{\}]
   \textcolor{OliveGreen}{\textbf{input}} \textcolor{blue}{\textbf{data}}                      \textcolor{teal}{\textit{( a single value: argument to append )}}
   \textcolor{OliveGreen}{\textbf{output}} \textcolor{blue}{\textbf{offsets0}} int32           \textcolor{teal}{\textit{( output Awkward Array offsets )}}
   \textcolor{OliveGreen}{\textbf{output}} \textcolor{blue}{\textbf{offsets1}} int32           \textcolor{teal}{\textit{( ... )}}
   \textcolor{OliveGreen}{\textbf{output}} \textcolor{blue}{\textbf{offsets2}} int32           \textcolor{teal}{\textit{( ... )}}
   \textcolor{OliveGreen}{\textbf{output}} \textcolor{blue}{\textbf{content}} float32          \textcolor{teal}{\textit{( and content )}}

   \textcolor{gray}{\textbf{0}} \textcolor{blue}{offsets0} \textcolor{OliveGreen}{\textbf{<-}} \textcolor{OliveGreen}{\textbf{stack}}             \textcolor{teal}{\textit{( offsets start at zero )}}
   \textcolor{gray}{\textbf{0}} \textcolor{blue}{offsets1} \textcolor{OliveGreen}{\textbf{<-}} \textcolor{OliveGreen}{\textbf{stack}}
   \textcolor{gray}{\textbf{0}} \textcolor{blue}{offsets2} \textcolor{OliveGreen}{\textbf{<-}} \textcolor{OliveGreen}{\textbf{stack}}

   \textcolor{OliveGreen}{\textbf{:}} \textcolor{blue}{\textbf{node3}}
     \textcolor{magenta}{\{float32-command\}} \textcolor{OliveGreen}{\textbf{=}} \textcolor{OliveGreen}{\textbf{if}}
       \textcolor{gray}{\textbf{0}} \textcolor{blue}{data} \textcolor{OliveGreen}{\textbf{seek}}                 \textcolor{teal}{\textit{( calling code puts the next value at )}}
       \textcolor{blue}{data} \textcolor{OliveGreen}{\textbf{d->}} \textcolor{blue}{content}            \textcolor{teal}{\textit{( the beginning of the input buffer )}}
     \textcolor{OliveGreen}{\textbf{else}}
       \textcolor{OliveGreen}{\textbf{halt}}                        \textcolor{teal}{\textit{( only the "float32" command is allowed )}}
     \textcolor{OliveGreen}{\textbf{then}}
   \textcolor{OliveGreen}{\textbf{;}}
   \textcolor{magenta}{\{node2\}} \textcolor{magenta}{\{node1\}} \textcolor{magenta}{\{node0\}}         \textcolor{teal}{\textit{( see below )}}

   \textcolor{gray}{\textbf{0}} \textcolor{OliveGreen}{\textbf{begin}}
     \textcolor{OliveGreen}{\textbf{pause}} \textcolor{blue}{node0}                   \textcolor{teal}{\textit{( pause for input, run forever )}}
   \textcolor{OliveGreen}{\textbf{again}}
\end{Verbatim}
\normalsize

\noindent The words in curly brackets are strings to be replaced, such as the following for \textcolor{magenta}{\tt \{nodeN\}}:

\small
\begin{Verbatim}[commandchars=\\\{\}]
   \textcolor{OliveGreen}{\textbf{:}} \textcolor{magenta}{\{node_name\}}
     \textcolor{magenta}{\{begin_list-command\}} \textcolor{OliveGreen}{\textbf{<>}} \textcolor{OliveGreen}{\textbf{if}}     \textcolor{teal}{\textit{( "begin_list" is required here )}}
       \textcolor{OliveGreen}{\textbf{halt}}
     \textcolor{OliveGreen}{\textbf{then}}
     \textcolor{gray}{\textbf{0}} \textcolor{OliveGreen}{\textbf{begin}}
       \textcolor{OliveGreen}{\textbf{pause}} \textcolor{OliveGreen}{\textbf{dup}} \textcolor{magenta}{\{end_list-command\}} \textcolor{OliveGreen}{\textbf{=}} \textcolor{OliveGreen}{\textbf{if}}
         \textcolor{OliveGreen}{\textbf{drop}}
         \textcolor{magenta}{\{offsets_name\}} \textcolor{OliveGreen}{\textbf{+<-}} \textcolor{OliveGreen}{\textbf{stack}}   \textcolor{teal}{\textit{( update offsets for this array node )}}
         \textcolor{OliveGreen}{\textbf{exit}}                       \textcolor{teal}{\textit{( exit this subroutine's infinite loop )}}
       \textcolor{OliveGreen}{\textbf{else}}
         \textcolor{magenta}{\{next_node_name\}}           \textcolor{teal}{\textit{( another list node like this or node3 )}}
         \textcolor{OliveGreen}{\textbf{1+}}
       \textcolor{OliveGreen}{\textbf{then}}
     \textcolor{OliveGreen}{\textbf{again}}
   \textcolor{OliveGreen}{\textbf{;}}
\end{Verbatim}
\normalsize

\noindent The \textcolor{magenta}{\tt \{*-command\}} substitutions are integers associated with each command. The virtual machine waits at a \textcolor{OliveGreen}{\tt\textbf{pause}} word until {\tt TypedArrayBuilder} puts a command number on the stack (and a number value in the input buffer for the {\tt float32} command), then resumes program flow, letting AwkwardForth format the output or \textcolor{OliveGreen}{\tt\textbf{halt}} if the wrong command is called. Thus, {\tt TypedArrayBuilder} can be statically compiled but ``wired'' to different actions at runtime.

\section{Performance}
\label{sec:performance}

\begin{figure}[p]
\includegraphics[width=\linewidth]{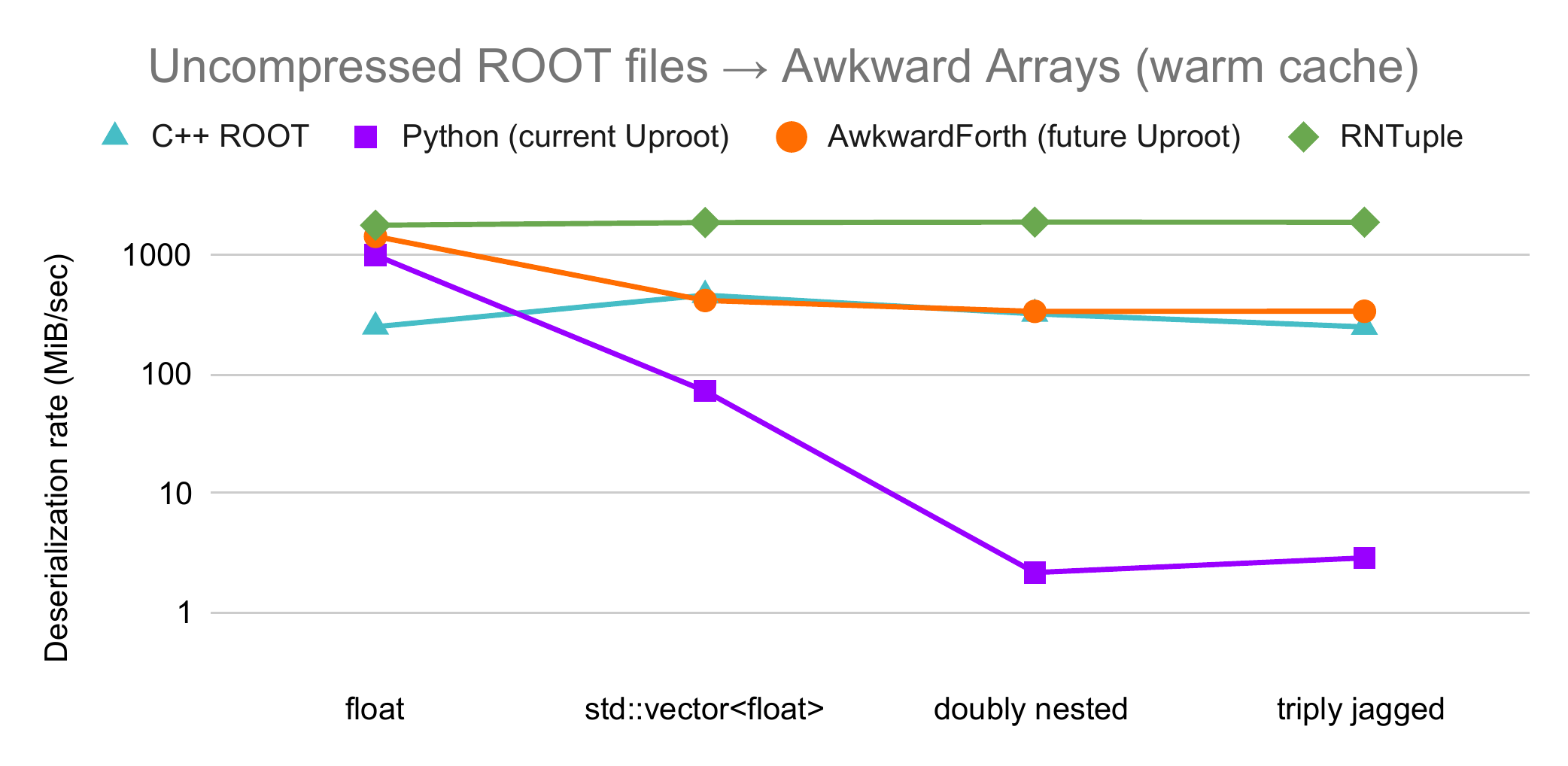}

\includegraphics[width=\linewidth]{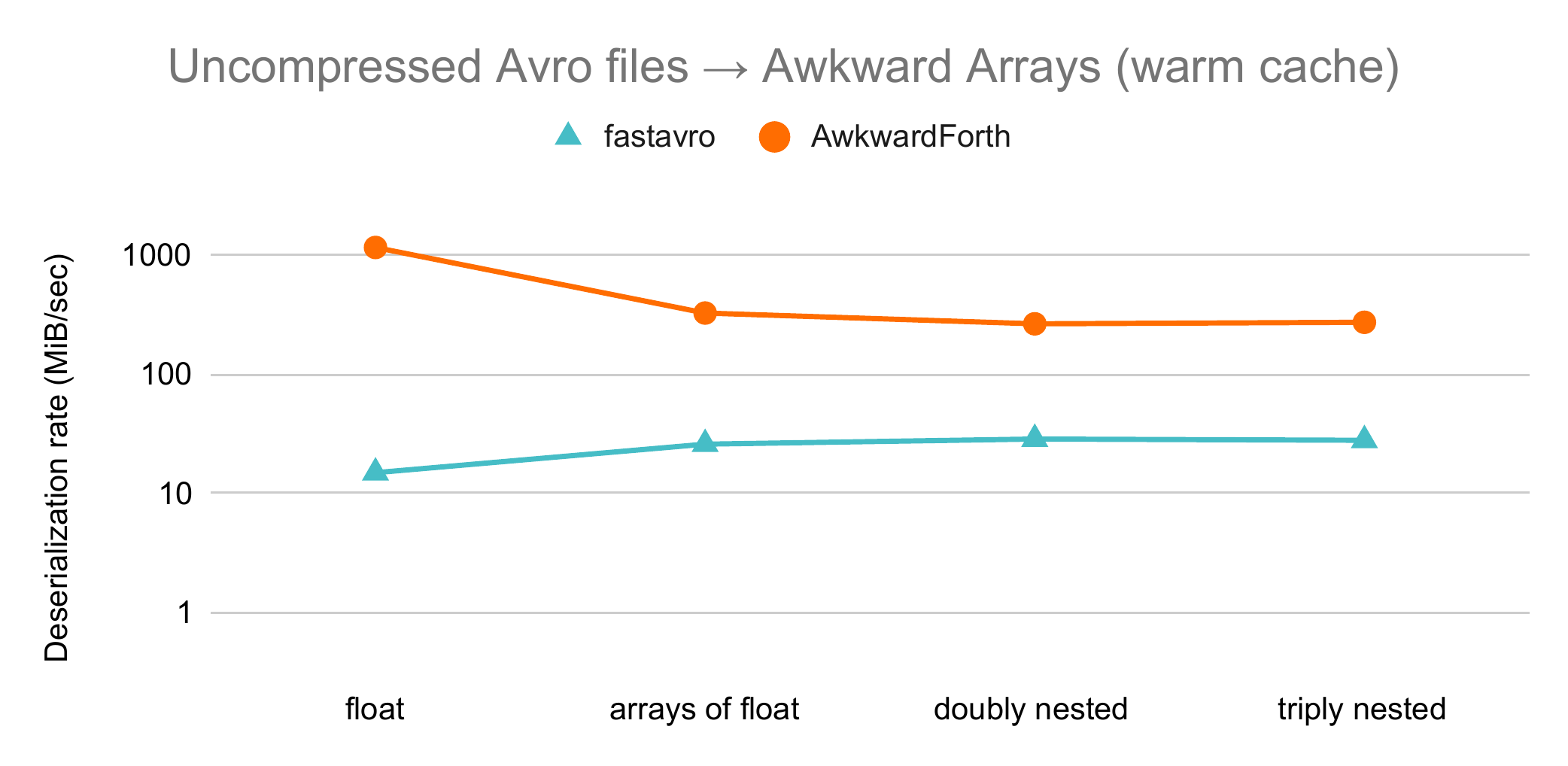}

\includegraphics[width=\linewidth]{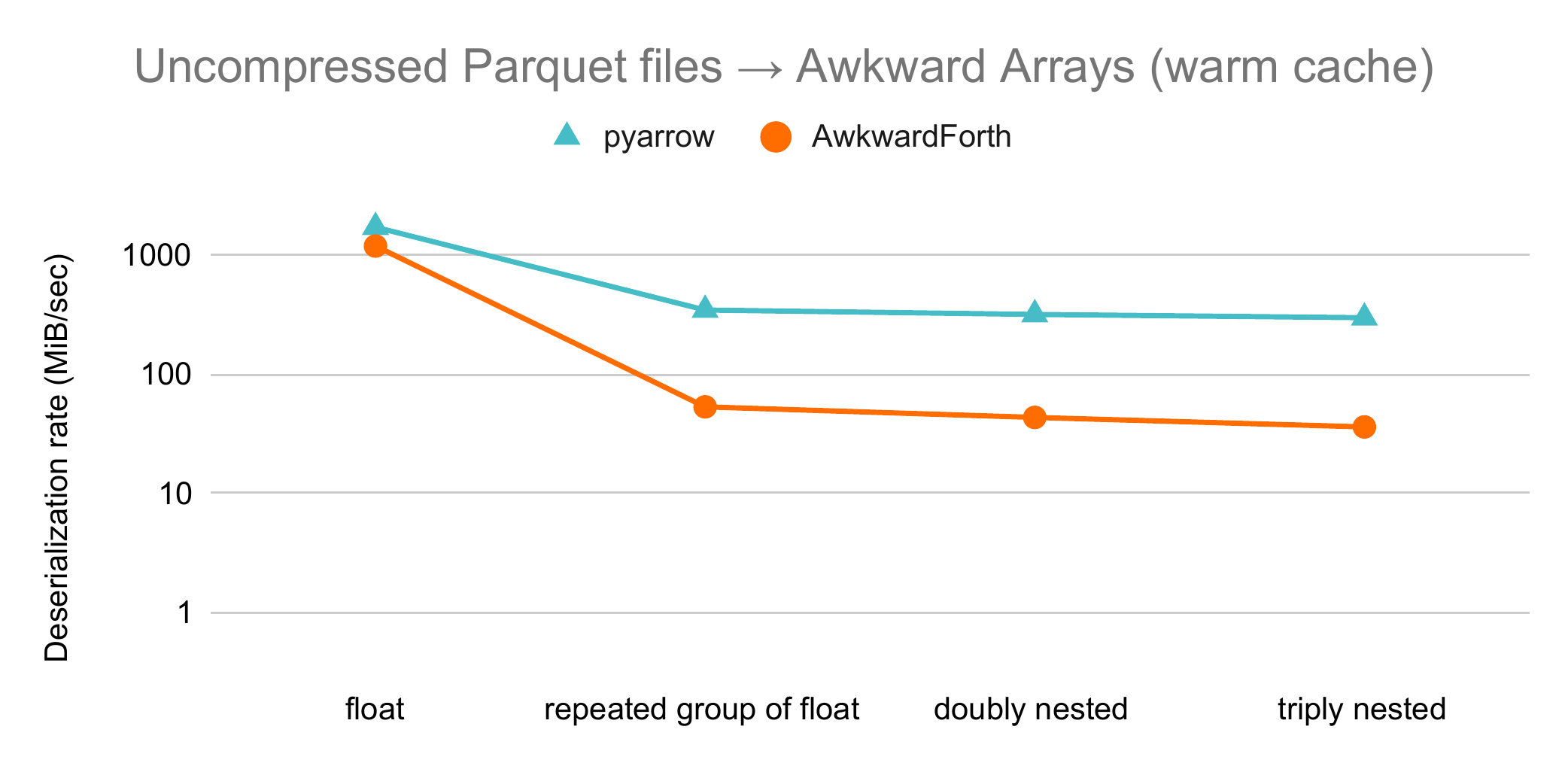}

\caption{Deserialization rate for converting uncompressed ROOT, Avro, and Parquet files to Awkward Arrays from warm cache. C++ ROOT, fastavro, and pyarrow are the leading libraries for each file type. \label{fig:performance}}
\end{figure}

Figure~\ref{fig:performance} presents single-threaded deserialization rates of uncompressed ROOT, Avro, and Parquet files from a warmed filesystem cache on an Intel i7-8750H (2.2~GHz) processor. There are four test files for each format: 4-byte \textcolor{Maroon}{\tt\textbf{float}}, variable-length lists of \textcolor{Maroon}{\tt\textbf{float}}, doubly nested lists of lists of \textcolor{Maroon}{\tt\textbf{float}}, and triply nested lists of lists of lists of \textcolor{Maroon}{\tt\textbf{float}}. For ROOT files, variable-length lists are {\tt std::vector}, but in Avro they are called ``arrays'' (despite being variable-length), and in Parquet, they are called ``repeated groups.'' All four types of files contain exactly 1\,073\,741\,824 \textcolor{Maroon}{\tt\textbf{float}} values and are all approximately 4~GiB (large, but well within the 15.5~GiB of physical RAM). The lengths of the lists at each level of depth are Poisson-distributed with a mean of 8.0 items per list. ROOT's TBasket size, Avro's data block size, and Parquet's row-group and page sizes were all fixed at 64~MiB.

The ROOT files were read in three ways: (1)~using C++ ROOT, loading data with {\tt TBranch::GetEntry}) and copying it into {\tt offsets0}, {\tt offsets1}, {\tt offsets2}, and {\tt content} buffers for Awkward Array, (2)~using Uproot 4.0.1's Python code, and (3)~using AwkwardForth (with Uproot to find the TBaskets within the file).

As an additional comparison, the same data were converted into ROOT's future RNTuple format, which is truly columnar. As expected (reproducing our previous results~\cite{chep2019}), \textcolor{Maroon}{\tt\textbf{float}} data are fastest to read, being essentially a memory-copy from TBaskets or RNTuple pages into the {\tt content} buffer. (ROOT's {\tt TBranch::GetEntry} is not as fast as direct memory access, which could be enabled by switching to ROOT's BulkIO feature, but we did not attempt that in this study.) RNTuple maintains this speed for all levels of nestedness for the same reason, while Uproot's NumPy tricks slow down reading of {\tt std::vector<\textcolor{Maroon}{\textbf{float}}>} and Uproot's pure Python is orders of magnitude slower for any deeper nesting. However, AwkwardForth keeps pace with ROOT's {\tt TBranch::GetEntry} at all levels of nestedness.

We next compared AwkwardForth with fastavro, the leading Python package for reading Avro files. fastavro is a Python extension library, using the C Avro implementation for speed. However, fastavro is compiled without knowledge of the schemas of the Avro files, which limits this advantage. AwkwardForth, on the other hand, has specialized Forth code for each data type and it's fast enough to read Avro 10--80$\times$ faster than fastavro.

Finally, we compared AwkwardForth with pyarrow, a Python extension library for the C++ Arrow and Parquet projects. In this case, pyarrow outperforms AwkwardForth by factors of 1.5--8$\times$. The AwkwardForth programs for Parquet in Section~\ref{sec:programs} show why: there's nothing about them that specializes to the data type except for the \textcolor{OliveGreen}{\tt\textbf{\#2bit->}} repetition level reader (deeper levels of nesting require logarithmically more bits) and the number of \textcolor{blue}{\tt count} variables and \textcolor{blue}{\tt offset} buffers. Any data type can be efficiently read with the same C++ code.

For compressed data (not shown in Figure~\ref{fig:performance}), all rates are suppressed by a constant for each algorithm. LZ4 is about 10\% slower than the memory copy, but ZLIB is 10$\times$ slower than the memory copy, so AwkwardForth is not the bottleneck for ZLIB-compressed data.

\begin{figure}[t]
\includegraphics[width=0.5\linewidth]{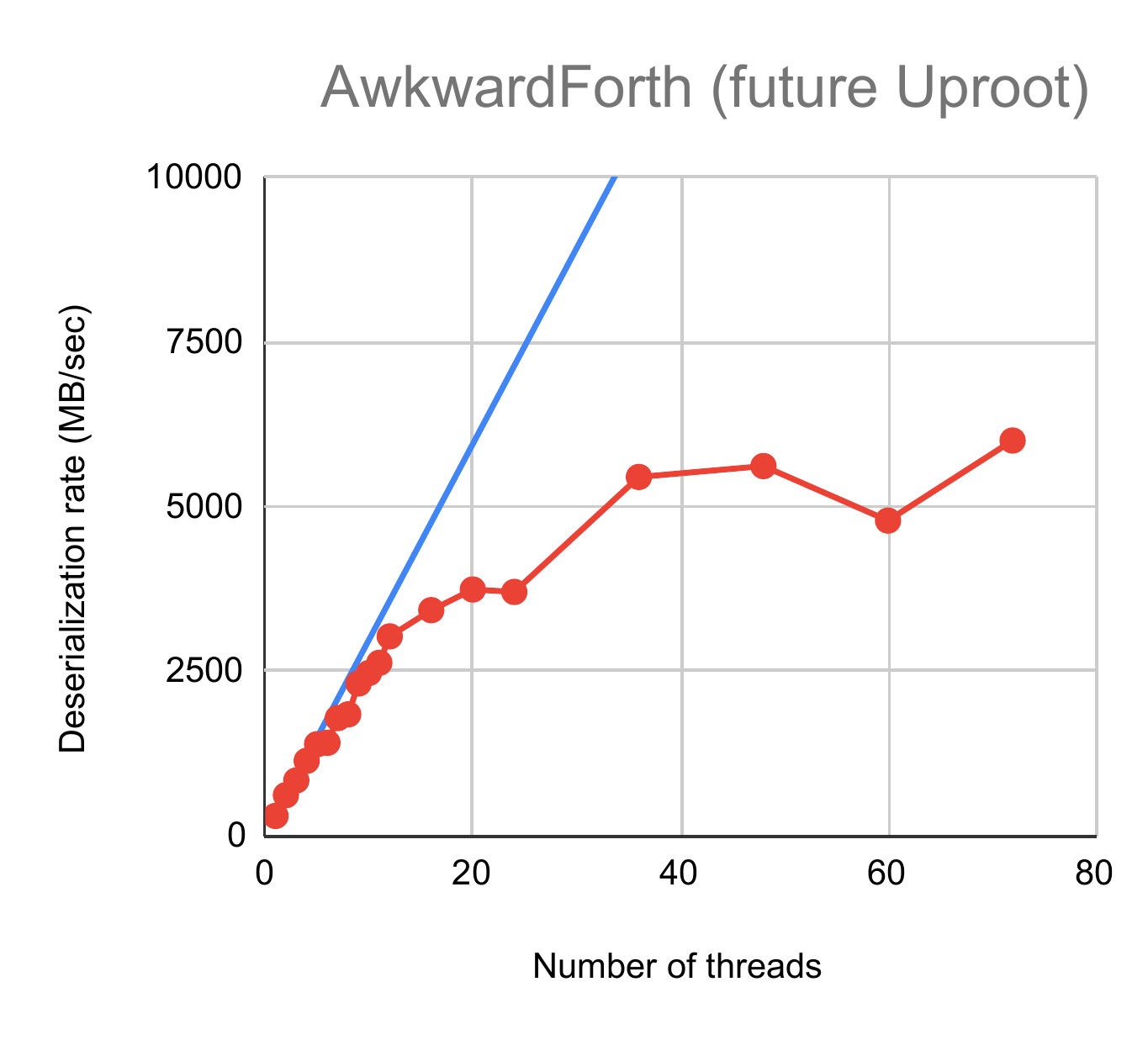}
\includegraphics[width=0.5\linewidth]{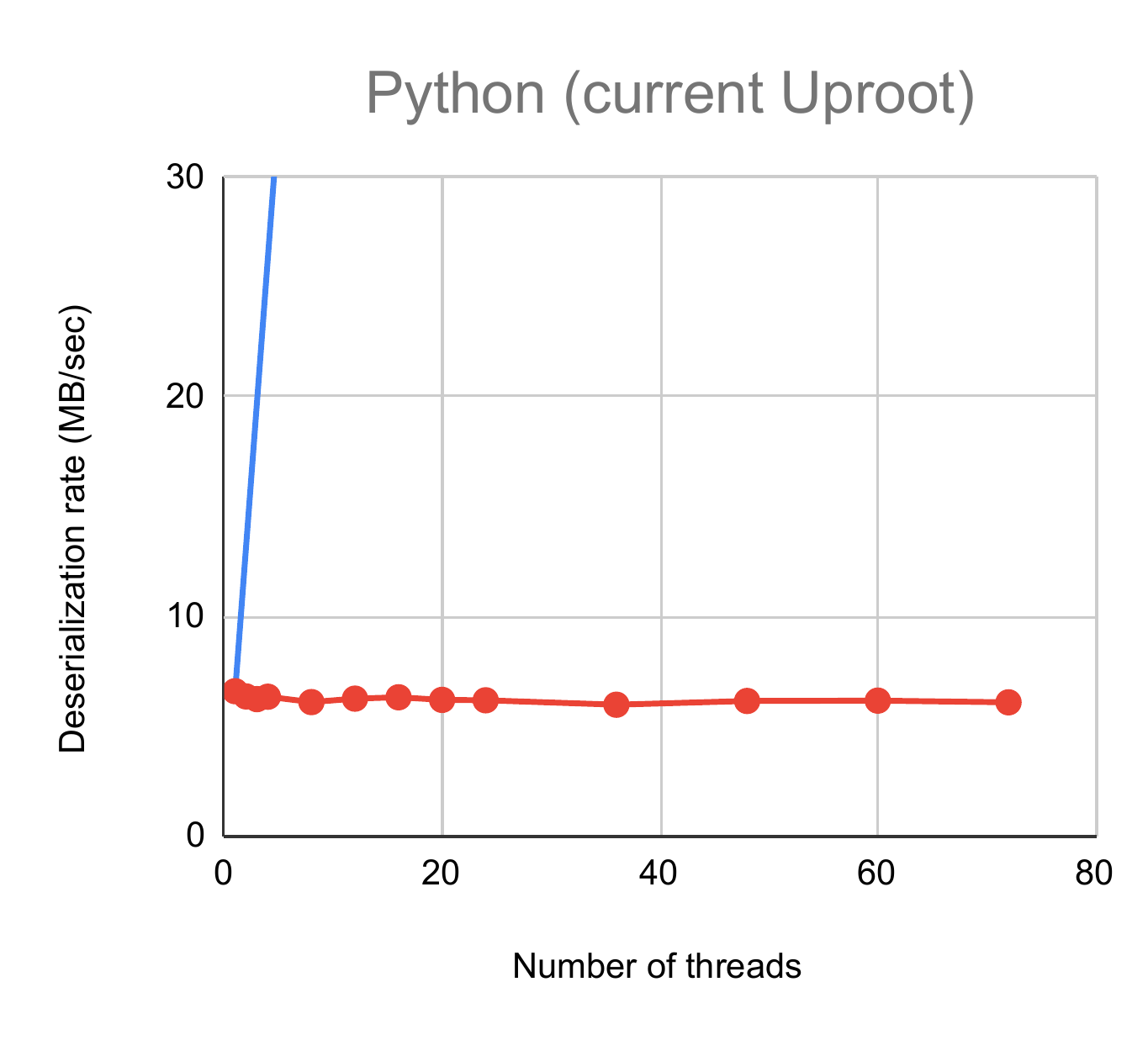}

\caption{Scaling of deserialization by the number of threads. AwkwardForth (when called from Python) releases the Python GIL, which otherwise prevents any gains from parallel processing. \label{fig:scaling}}
\end{figure}

We also studied the scaling of AwkwardForth with threads. The AwkwardForth virtual machine is stateful, but lightweight: many can be launched at once, one for each thread. Figure~\ref{fig:scaling} shows deserialization rate of the ROOT files on an AWS {\tt c5.18xlarge} instance, which has 72~cores. Linear scaling falls off at about 20~threads, or 4--5~GiB per second, which may be a limitation of memory bus supplying data from RAM. Python, by comparison, does not scale at all because of its Global Interpreter Lock (GIL)~\cite{python-gil}.

All scripts used to produce Figures~\ref{fig:performance} and \ref{fig:scaling} are available on GitHub~\cite{this-study}.

\section{Conclusions}

AwkwardForth is a ``lightweight'' solution to the problem of generating fast deserialization code whose form is only known at runtime. In all, it consists of 3388 lines of C++ code in Awkward Array (version~1.1.0), with no dependencies. Unlike user-facing DSLs, its syntax is not subject to usability constraints, only ease of algorithmic generation and speed. As shown in Section~\ref{sec:performance}, it keeps pace with ROOT's own TTree deserialization, exceeds fastavro for Avro, and is slower than but within an order of magnitude of pyarrow for Parquet.

Most importantly, AwkwardForth vastly improves upon the Python code in Uproot, speeding up {\tt std::vector<std::vector<\textcolor{Maroon}{\textbf{float}}>{}>} deserialization by factors of hundreds, and it makes parallel TBasket reading worthwhile. It will be integrated into Uproot this year (2021).

\section{Acknowledgements}

This work was supported by the National Science Foundation under Cooperative Agreement OAC-1836650 (IRIS-HEP).

\bibliography{bibfile}

\begin{thebibliography}{19}

\bibitem{uproot}
J.~Pivarski et~al., \emph{{Uproot}},
  \urlstyle{tt}\url{https://doi.org/10.5281/zenodo.4543730}

\bibitem{root}
R.~Brun, F.~Rademakers, \emph{{ROOT: An object oriented data analysis
  framework}}, Nucl. Instrum. Meth. A \textbf{389}, 81 (1997)

\bibitem{numpy}
C.R. Harris, K.J. Millman, S.J. van~der Walt, R.~Gommers, P.~Virtanen,
  D.~Cournapeau, E.~Wieser, J.~Taylor, S.~Berg, N.J. Smith et~al., \emph{Array
  programming with {NumPy}}, Nature \textbf{585}, 357 (2020)

\bibitem{awkward}
J.~Pivarski et~al., \emph{{Awkward Array}},
  \urlstyle{tt}\url{https://doi.org/10.5281/zenodo.4539721}

\bibitem{chep2019}
J.~{Pivarski}, P.~{Elmer}, D.~{Lange}, \emph{{Awkward Arrays in Python, C++,
  and Numba}}, in \emph{European Physical Journal Web of Conferences} (2020),
  Vol. 245 of \emph{European Physical Journal Web of Conferences}, p. 05023,
  \texttt{2001.06307}

\bibitem{cling}
V.~Vasilev, P.~Canal, A.~Naumann, P.~Russo, \emph{Cling {\textendash} The New
  Interactive Interpreter for {ROOT} 6}, Journal of Physics: Conference Series
  \textbf{396}, 052071 (2012)

\bibitem{postscript}
G.~Reid, \emph{Thinking in PostScript} (Addison-Wesley, 1990), ISBN
  9780201523720,
  \urlstyle{tt}\url{https://books.google.com/books?id=ZeRXvgAACAAJ}

\bibitem{forth-standard}
{Technical Committee X3J14}, \emph{{ANSI Forth Standard}} (1994),
  \urlstyle{tt}\url{https://www.taygeta.com/forth/dpans.html}

\bibitem{forth}
E.D. Rather, D.R. Colburn, C.H. Moore, \emph{The Evolution of Forth}, in
  \emph{The Second ACM SIGPLAN Conference on History of Programming Languages}
  (Association for Computing Machinery, New York, NY, USA, 1993), HOPL-II, p.
  177–199, ISBN 0897915704,
  \urlstyle{tt}\url{https://doi.org/10.1145/154766.155369}

\bibitem{protobuf}
\emph{{Google ProtoBuf}},
  \urlstyle{tt}\url{https://developers.google.com/protocol-buffers/}

\bibitem{thrift}
\emph{{Apache Thrift}}, \urlstyle{tt}\url{https://thrift.apache.org/}

\bibitem{avro}
\emph{{Apache Avro}}, \urlstyle{tt}\url{http://avro.apache.org/}

\bibitem{flatbuffers}
\emph{{Google FlatBuffers}},
  \urlstyle{tt}\url{https://google.github.io/flatbuffers/}

\bibitem{json}
\emph{{JSON}}, \urlstyle{tt}\url{https://www.json.org/}

\bibitem{rntuple}
J.~{Blomer}, P.~{Canal}, A.~{Naumann}, D.~{Piparo}, \emph{{Evolution of the
  ROOT Tree I/O}}, in \emph{European Physical Journal Web of Conferences}
  (2020), Vol. 245 of \emph{European Physical Journal Web of Conferences}, p.
  02030, \texttt{2003.07669}

\bibitem{parquet}
\emph{{Apache Parquet}}, \urlstyle{tt}\url{https://parquet.apache.org/}

\bibitem{arrow}
\emph{{Apache Arrow}}, \urlstyle{tt}\url{https://arrow.apache.org/}

\bibitem{python-gil}
D.~Beazley, \emph{{Understanding the Python GIL}}, in \emph{PyCON Python
  Conference. Atlanta, Georgia} (2010)

\bibitem{this-study}
\emph{{Performance measurement code for this study}},
  \urlstyle{tt}\url{https://github.com/scikit-hep/awkward-1.0/tree/1.1.0/studies/awkward-forth-performance}

\end{thebibliography}





\end{document}